\documentclass{ws-ijmpd}
\usepackage[super,compress]{cite}
\usepackage{graphicx}
\begin{document}
\markboth{C. Deliduman \& B. Yap\i\c{s}kan}{Absence of Relativistic Stars in $f(T)$ Gravity}

\renewcommand{\thefootnote}{\fnsymbol{footnote}}

\newcommand{\fr}{\frac}
\newcommand{\ct}{\cite}
\newcommand{\lb}{\label}
\newcommand{\ti}{\tilde}
\newcommand{\prm}{\prime}

\newcommand{\al}{\alpha}
\newcommand{\bt}{\beta}
\newcommand{\ka}{\kappa}
\newcommand{\la}{\lambda}
\newcommand{\La}{\Lambda}
\newcommand{\si}{\sigma}
\newcommand{\Si}{\Sigma}
\newcommand{\te}{\theta}
\newcommand{\Te}{\Theta}
\newcommand{\ga}{\gamma}
\newcommand{\ep}{\epsilon}
\newcommand{\ups}{\upsilon}
\newcommand{\Om}{\Omega}
\newcommand{\om}{\omega}
\newcommand{\ld}{\delta}
\newcommand{\CD}{\Delta}
\newcommand{\ot}{\otimes}
\newcommand{\vae}{\varepsilon}
\newcommand{\vap}{\varphi}

\newcommand{\app}{\approx}
\newcommand{\ra}{\rightarrow}
\newcommand{\eqv}{\equiv}
\newcommand{\di}{\diamond}
\newcommand{\tbf}{\textbf}
\newcommand{\mbf}{\mathbf}
\newcommand{\oln}{\overline}
\newcommand{\dgr}{\dagger}
\newcommand{\del}{\partial}

\newcommand{\sqr}{\square}
\newcommand{\nno}{\nonumber}
\newcommand{\noin}{\noindent}
\newcommand{\bit}{\bibitem}

\def\ds{\displaystyle}

%
\catchline{}{}{}{}{}
%

\title{Absence of Relativistic Stars in $f(T)$ Gravity}

\author{Cemsinan Deliduman$^\dag$ and Bar\i \c{s} Yap\i\c{s}kan$^\ddag$}

\address{Department of Physics, Mimar Sinan Fine Arts University,
Bomonti 34380, \.{I}stanbul, Turkey\\
$^\dag$cemsinan@msgsu.edu.tr\\ 
$^\ddag$baris.yapiskan@msgsu.edu.tr}

\maketitle

\begin{history}
\end{history}

\begin{abstract}
In this work we investigate the structure of neutron stars in
modified $f(T)$ gravity models. We find that, unlike the $f(R)$ models, the equations of motion
put a rather strict constraint on the possible $f(T)$ functions.
Specifically, after analyzing the problem in two different choice of coordinates with
spherical symmetry, we conclude that the relativistic neutron star solution in
$f(T)$ gravity models is possible only if $f(T)$ is a
linear function of the torsion scalar $T$, that is in the case of
Teleparallel Equivalent of General Relativity.

\end{abstract}

\ccode{PACS numbers: 04.50.Kd, 97.60.Jd}

\tableofcontents


\section{Introduction}

One of the assumptions of the current paradigm of cosmology is the
validity of Einstein's theory of gravity in all scales: from
phenomena we observe in our solar system, to the large scale
structure of the universe. However, in recent years, data from
distant supernovae Ia \cite{Perlmutter,Riess1,Riess2,Kowalski} are
interpreted as evidence of late time acceleration in the expansion rate
of the universe. Continuing to assume the validity of general
relativity in all scales and best fit to observational data
requires existence of a non--vanishing positive cosmological
constant. There are several theoretical problems related with
the existence of cosmological constant (see for example
\cite{Weinberg,Peebles,Nobbenhuis,Bousso}), most important of which
is the lack of a quantum theoretical method to calculate its
inferred value from cosmological data. 
Therefore, several authors tried to avoid the problems of
cosmological constant with alternative routes of explanations.
To explain the late time accelerated expansion one
can either modify ``wood" part, i.e. matter part, or modify
``marble" part, i.e. geometric part, of the Einstein's field
equations. In the former approach one adds the energy--momentum
component of dark energy with an equation of state $p/\rho \approx
-1$ to the wood part of the Einstein's equations. As opposed to this, in
the latter approach, one modifies the theory of gravity, and this
way changes the marble part of the Einstein's equations. Such
modifications could be in two fashions: one can either increase the
number of degrees of freedom by adding new gravitational fields into
the theory, or changes the form of the gravity action without introducing
new fields. Metric or vierbein field still remains the only
gravitational degree of freedom in the second approach. This is the approach we are going 
to take in this paper. There is also an approach in which one starts by modifying 
the gravity theory, and then moving all the modifications in the marble part to the wood
part, one interprets these modifications as the energy--momentum tensor of dark 
energy. Such line of reasoning will not be followed in the present article.

One important family of modifications of Einstein--Hilbert action is
the $f(R)$ theories of gravity (see \cite{Odintsov-rev,Capozziello-rev,Sotiriou-rev,deFelice-rev}
and references therein). In such theories one uses a function of curvature scalar as the
Lagrangian density. However, the field equations of $f(R)$ gravity
models turn out to be fourth order differential equations in the metric
formalism and therefore they are difficult to analyze. With a
similar line of thought one can also modify teleparallel equivalent
of general relativity \footnote[4]{We will sometime use the shorter
name, ``teleparallel gravity," instead of the longer name ``teleparallel
equivalent of general relativity" to describe the same theory.}.
This theory is defined on a Witzenb\"{o}ck space--time, which is
curvatureless, but has a non--vanishing torsion. Lagrangian density
is equivalent to the torsion scalar and the field equations of
teleparallel gravity are exactly the same as the Einstein's equations in
any background metric \cite{Einstein,Unzicker,Hayashi, Andrade}.
One can modify teleparallel gravity by having
a Lagrangian density equivalent to a function of torsion scalar. This is first done in the context
of Born--Infeld gravity \cite{Ferraro1,Ferraro2}, however it is possible to have any function
$f(T)$. Then one has $f(T)$ theories of gravity \cite{Bengochea}. These theories are more
manageable compared to $f(R)$ theories, because their field
equations are second order differential equations.

A modified gravity theory should be able to pass several tests
before it can be considered a viable theory of gravity. In the weak
gravity regime, such a theory should  be compatible with solar system
tests and table--top experiments. In cosmological scales, it should
produce late time acceleration, be free of gravitational
instabilities, and obey constraints of standard model of cosmology.
Such a theory should do well also in strong gravity regime, for
example it should have solutions of neutron stars with mass--radius
relation not conflicting with the current observations. In this paper we
are analyzing $f(T)$ gravity in the strong field regime and test
whether such theories could be viable theories of reality. To our
surprise, we find that the ``non--diagonal'' field equations of $f(T)$ gravity theories
require $f(T)$ to be a linear function of torsion scalar $T$, otherwise there are no
solutions for relativistic neutron stars in such a theory. This
means that only teleparallel equivalent of general relativity could
be a viable theory of gravity. Any modification of it, other than addition
of a cosmological constant, will not have relativistic neutron star solution and
therefore be at odds with the observations.

The plan of this paper is as follows. In the next section we will
summarize main aspects and provide the field equations of teleparallel
and $f(T)$ theories of gravity. Then, in section (\ref{contra}), the
field equations will be rewritten in a spherically symmetric
background with a diagonal metric. This is the metric seen by a stationary
observer far away from the neutron star.
Since $f(T)$ gravity theories are not Lorentz invariant \cite{Sot1,Sot2},
we are going to repeat our analysis also for a non-diagonal spherically symmetric
metric in section (\ref{GP}). This other metric is the Gullstrand--Painlev\'e
type and it is the metric seen by a free falling observer.
We are going to discuss implications of the main result of this paper in the conclusions. 


\section{Field Equations of $f(T)$ Gravity}

As is well known, general relativity is formulated on a
pseudo--Riemannian manifold and its dynamical variable is the metric
tensor defined on that manifold. Through metric tensor one defines
the Levi--Civita connection, and then Riemann and the related
tensors. Torsion tensor in a Riemannian space--time vanishes due to
the symmetry properties of the Levi--Civita connection. 
If the connection is different than the Levi--Civita connection, then 
torsion tensor is non-zero together with the Riemann tensor, then we
have Riemann--Cartan space--time on which Einstein--Cartan theory of
gravity is defined. In a sense, Riemannian space--time can be thought
of a subclass of Riemann--Cartan space--time: by setting the torsion
tensor to zero, Riemann--Cartan space--time is reduced to a
Riemannian space--time. Another subclass is obtained by setting the curvature tensor to
zero instead. Then affine connections are no longer symmetric and while the Riemann tensor, which is a measure of curvedness of the space-time manifold, vanishes everywhere, the torsion tensor is non-zero. This results in the possibility of defining a distinguished vector field, which points the same direction at each point of the space-time manifold, hence the property of teleparallelizability. This is the so called Wietzenb\"{o}ck space--time \cite{W}, and the theory of gravity defined on it is called teleparallel gravity.

Dynamical field of teleparallel gravity is the vierbein field
$e^i_{\mu}$ which is given in terms of the metric tensor as
\begin{equation}
g_{\mu \nu} =  \eta_{ij} e^i_{\mu} e^j_{\nu},
\end{equation}
where latin indices label coordinates of tangent space  and
greek indices label coordinates of the space--time. Both set of
indices run over $(0,1,2,3)$. Teleparallel gravity differ from the
general relativity in the sense that it uses the curvatureless
Weitzenb\"{o}ck connection
\begin{equation} \lb{wiet}
\Gamma^{\rho}_{\;\;\mu \nu} = e_i^{\rho} \partial_{\nu}e^i_{\mu} .
\end{equation}
Then non--vanishing torsion is given in terms of Wietzenb\"{o}ck connection as
 \begin{equation}
T^\rho_{\;\;\mu\nu}= \Gamma^{\rho}_{\;\;\nu \mu} - \Gamma^{\rho}_{\;\;\mu \nu}
=e_i^\rho(\del_\mu e_\nu^i - \del_\nu e_\mu^i) .
 \end{equation}
To define the action for teleparallel gravity one defines two more tensors:
one is the contorsion tensor given in terms of the torsion tensor,
\begin{equation}
K^{\mu\nu}_{\quad\rho}=-\frac{1}{2}(T^{\mu\nu}_{\quad\rho}-T^{\nu\mu}_{\quad\rho}
-T_{\rho}^{\;\;\mu\nu}) ,
\end{equation}
and the other tensor is defined in terms of contorsion and torsion tensors as
\begin{equation}
S_\rho^{\;\;\mu\nu}=\frac{1}{2}(K^{\mu\nu}_{\quad\rho}+\ld^\mu_\rho
T^{\al\nu}_{\quad\al}-\ld^\nu_\rho T^{\al\mu}_{\quad\al}) .
\end{equation}
Then the torsion scalar is defined to be
\begin{equation} \lb{trsc}
T=S_\rho^{\;\;\mu\nu}T^\rho_{\;\;\mu\nu} .
\end{equation}

Torsion scalar is used as the Lagrangian density in the action for teleparallel gravity,
\begin{equation}
S=-\fr{1}{16\pi G}\int d^4 x\, e\, T + S_{\rm{matter}}\, ,
\end{equation}
where $e=det(e^i_\mu)=\sqrt{-g}$ and $S_{\rm{matter}}$ is the part
of the action that describes matter fields interacting with the
vierbein field. The variation of the action with respect to the
vierbein leads to the equations of motion
which are identical to the equations of motion of general relativity. 
Therefore this form of teleparallel
gravity is equivalent to the general relativity. This theory is
first proposed by Einstein \cite{Einstein,Unzicker} and therefore it is rightfully called
Einstein's other gravity \cite{Linder}, new general relativity \cite{Hayashi} or
teleparallel equivalent of general relativity \cite{Andrade}.

As in the case of general relativity, this action might be modified by having a function of torsion scalar,
$f(T)$, as the Lagrange density,
\begin{equation}
\label{act}
S=-\fr{1}{16\pi G}\int d^4 x\, e\, f(T) + S_{\rm{matter}} .
\end{equation}
Then we have $f(T)$ theories of gravity similar to $f(R)$ theories
of gravity. As stated in the introduction, this is a new set of
modified gravity theories which might have the potential to answer
some unresolved questions in the contemporary cosmology. The
variation of the action for $f(T)$ gravity with respect to the
vierbein leads to the following field equations,
\begin{equation}
e_i^\rho S_\rho^{\;\;\mu\nu}\del_\mu T f_{TT}
+ e^{-1}\del_\mu(e\ e_i^\rho\ S_\rho^{\;\;\mu\nu})f_T 
+ e_i^\mu T^\la_{\;\;\mu\ka}S_\la^{\;\;\nu\ka}f_T
- \frac{1}{4}e_i^\nu f =-4\pi e_i^{\la}T_\la^\nu \, ,  \label{meom}
\end{equation}
where $T_{\mu\nu}$ is the energy-momentum tensor of the particular matter, whereas $f_T$
and $f_{TT}$ represent first and second derivatives of $f(T)$ with respect to the
torsion scalar $T$, respectively.
Note that we are setting $c=1$ and $G=1$ here and for the rest of this paper.


\section{Reference Frame of a Distant Stationary Observer} \lb{contra}

To describe relativistic neutron stars in a theory of gravity, one usually starts with
two assumptions: 1) the spherically symmetric metric of neutron star has diagonal structure,
\begin{equation}
\label{met}
ds^2=-e^{2\Si(r)}dt^2 + e^{2\La(r)}dr^2 + r^2 d\te^2 + r^2 \sin^2\te d\phi^2 ,
\end{equation}
and 2) matter inside the neutron star is a perfect fluid which has a
diagonal energy--momentum tensor in the rest frame of the matter,
\begin{equation}
T_\mu^{\; \nu}=diag(-\rho,p,p,p)
\end{equation}
where $\rho$ and $p$ are the energy density and pressure of the
fluid, respectively. Matter functions, $\rho$ and $p$, and metric
functions, $\Si$ and $\La$, are taken independent of time, which
means that the system is in equilibrium, and due to spherical symmetry
they are functions of $r$ only.

\textit{Physically,} (\ref{met}) is the metric seen by a distant observer. If we were solving for the vacuum field equations in general relativity we would get the Schwarzschild metric as the solution outside a spherical star. Then our outside observer would observe existence of event horizon, even though an infalling observer would not observe such a special place. In the case of neuron star the outside solution is assumed to be still given by the Schwarzschild metric and Tolman--Openheimer--Volkov (TOV) equations give us the internal solution, which is matched with the Schwarzschild solution at the surface of the star.

Energy--momentum tensor is taken to be covariantly constant in general relativity, which leads to
\begin{equation} \lb{cons}
\frac{dp}{dr}=-(\rho+p)\frac{d\Si}{dr} .
\end{equation}
This equation should hold identically for all systems described with the metric (\ref{met}). This is an
equation of hydrostatic equilibrium and global aspects of neutron
stars, such as mass--radius relation, can be determined from it, if
the metric function $\Si(r)$ is known. For a spherically symmetric
object in general relativity, this is one of the TOV equations after $\Si(r)$
is solved from the field equations.

The vierbein field $e^i_{\mu}$ derived from the metric (\ref{met}) is
\footnote{For clarity, we distinguish tangent space indices from space-time indices using parentheses.}
\begin{equation}
e^{(0)}_0=e^{\Si(r)}\ ,\quad e^{(1)}_1=e^{\La(r)}\ ,\quad 
e^{(2)}_2=r\ ,\quad e^{(3)}_3=r\sin \te .
\end{equation}
With these values, the determinant of vierbein becomes
$e=\sqrt{-g}=r^2\sin \te e^{(\Si+\La)}$, and the torsion scalar
(\ref{trsc}) in this background is found to be
\begin{equation}
T=-\fr{2}{r}(2\Si'+\fr{1}{r})e^{-2\La},
\end{equation}
where prime denotes differentiation with respect to $r$.

Substituting these values into the field equations of modified gravity (\ref{meom}), we obtain modified
equations of motion for $i=\nu=0$:
\begin{equation} \lb{mschw1}
16\pi \rho = -\frac4r e^{-2\La}T' f_{TT} +(\frac{2}{r^2}+2T+\frac4r (\Si'+\La')e^{-2\La})f_T - f\, ,
\end{equation}
and for $i=\nu=1$:
\begin{equation} \lb{mschw2}
16\pi p=-(\fr{2}{r^2}+2T)f_T + f .
\end{equation}
The other two ``diagonal'' components of the field equations are the same and are given as
\begin{equation} \lb{mschw3}
16\pi p = 2e^{-2\La} (\Si''+\Si'(\Si'-\La')+\frac1r (3\Si'-\La')+\frac{1}{r^2})f_T 
+ 2(\Si'+\frac{1}{r})e^{-2\La}T' f_{TT} + f .
\end{equation}
One can check that these equations become the same as the field equations 
of general relativity in the limit that $f(T)=T$.
Unlike the case of general relativity, in this theory we have an extra ``non--diagonal'' equation, which is obtained for $i=1$ and $\nu=2$ as
\begin{equation}
T' f_{TT} =0 .
\end{equation}

This equation puts a rather strict constraint on the possible $f(T)$ functions. In fact, the only 
possible $f(T)$ function is the linear one. As a result, among the
$f(T)$ theories of gravity, only in the case of teleparallel equivalent of general
relativity solutions for relativistic neutron stars exist.  One might try to argue that one
does not need to set $f_{TT}=0$ as is done here, but instead set $T'=0$ in which case one could
consider a general $f(T)$ gravity theory with an everywhere constant torsion scalar, $T=T_0$. However such a solution would not make sense as a relativistic star solution. This can be easily seen if one inspects the equation (\ref{mschw2}). For a constant $T$, the right hand side of this
equation blows up as $r \rightarrow 0$. This means that matter's pressure would also blow up. Such a solution cannot be considered to correspond to a physical neutron star. Therefore setting 
the torsion scalar to a constant value has some ill consequences as described. The plausible choice is to set $f_{TT}=0$, in which case no $f(T)$ gravity models other than the teleparallel equivalent of general relativity would be possible.

We would like to emphasize that the claim made above is concerned solely with the neutron star solution in $f(T)$ theories of gravity. In such a solution, it is expected that the energy density and pressure to have non-zero finite values only up to the radius of the neutron star. There could be some exotic spherically symmetric solutions of $f(T)$ gravity for which this is not the case: either pressure or energy density have everywhere non-zero values, or one or both take infinite value at some point. Even if such solutions could exist, the existence of them do not contradict the conclusions of this paper. We do not claim that no spherical solutions exists, but no neutron star solution exists in the sense of Tolman, Oppenheimer and Volkov.

One might argue against the conclusion of this section on the grounds that the $f(T)$ theories of
gravity lack Lorentz invariance \cite{Sot1,Sot2} and therefore result of this section could be just the artifact of the frame used. In an another frame we might not find the same result and relativistic neutron star solution could be natural there. If that is so, then we have very strong frame dependent physics: in one frame we have a universe without neutron stars and in the other with neutron stars. Such a frame dependent physics seems most bizarre. We are against this line of thought. As it is commented in \cite{Sot1}, because of absence of local Lorentz invariance one cannot choose vierbein field, but has to determine it from the field equations. However, this is very complicated computationally. 

Even though we might not be able to check the possibility of relativistic neutron star solution in all choices of frame corresponding to spherically symmetric background and find if relativistic star solution is possible in one of them, we would like to argue that there are \textit{two important physical observers,} whose observations should be enough to decide about the issue \textit{physically}. Note that the diagonal metric we utilized in this section is the metric seen by a distant stationary observer. This observer concludes that there are no neutron star solutions. In order to check this conclusion we could ask the same question to one other observer who is in a reference frame which is free falling to the neutron star. The Gullstrand--Painlev\'e type metric that we are going to use in the next section is the metric seen by such an observer. We are going to see that the freely falling observer's conclusion turns out to be the same as the stationary observer. These are the two important physical observers. If both of these observers cannot observe existence of neutron stars then observations of an observer with an obscure frame would not mean much \textit{physically}. 

A metric similar to Gullstrand--Painlev\'e metric is also used in \cite{Wang}. There, it is firstly shown that, in the case of Robertson--Walker metric, depending on the choice of frame, one obtains sets of field equations that give contradictory physics. Then, the generalized Gullstrand--Painlev\'e metric is used to search spherically symmetric static solutions. Curiously, in the mentioned coordinate system \cite{Wang}, it is shown that the Reissner--Nordstr\"om solution does not exist in $f(T)$ gravity theories other than the teleparallel equivalent of general relativity. This result is similar to our result about the absence of relativistic neutron stars in $f(T)$ theories of gravity.


\section{Reference Frame of a Freely Falling Observer} \lb{GP}

To see if the result of the previous section still holds in the case of a non--diagonal spherically
symmetric metric, we now work with a Gullstrand-Painlev\'e type metric
\begin{equation}
\lb{GPm}
ds^2=-\bt^2 dt^2+\ld_{ab}(\bt\sqrt{\al}\fr{x^a}{r}dt+dx^a)(\bt\sqrt{\al}\fr{x^b}{r}dt+dx^b)\, ,
\end{equation}
where latin indices from the beginning of alphabet run over $(1,2,3)$ and $r=\sqrt{x^a x_a}$ is the radial coordinate. Metric functions, $\alpha$ and $\beta$, depend on $r$ only. Time
coordinate of this metric is conformally Cartesian and space coordinates $x^a$ fully span the range
$(-\infty, +\infty)$. In order see the relation of this metric with (\ref{met}) we write it in terms of spherical coordinates:
\begin{equation}
\label{ GPm1}
ds^2 = -\beta^2 (1-\alpha)dt^2 + 2\beta\sqrt{\alpha}dt dr + dr^2 + r^2 d\Omega^2 .
\end{equation}
This form of metric is related to a diagonal form of metric via a Lorentz transformation
done only in the time direction:
\begin{equation}
\label{Lorentz}
\beta dt = \beta d\tilde{t}+ \frac{\sqrt{\alpha}}{1-\alpha}dr .
\end{equation}
With this transformation we obtain a more familiar form of the metric
\begin{equation}
\label{GPm2}
ds^2 = -\beta^2 (1-\alpha)d\tilde{t}^2 + \frac{1}{1-\alpha}dr^2 + r^2 d\Omega^2 .
\end{equation}

The vierbein field $e_{\mu}^i$ derived from the metric (\ref{GPm}) is
\begin{equation}
e_0^{(0)}=\bt, \qquad e_0^{(a)}=\bt\sqrt{\al}\fr{x^a}{r}, \qquad e_a^{(b)}=\ld_a^b .
\end{equation}
The determinant of the vierbein field is $e=\sqrt{-g}=\bt$ and torsion scalar is found equal to
\begin{equation} \label{T}
T=\fr{2\al}{r}(\fr{2\bt'}{\bt}+\fr{\al'}{\al}+\fr{1}{r}) .
\end{equation}
Substituting these values into the field equations of $f(T)$ gravity (\ref{meom}) and using again the energy--momentum tensor of perfect fluid, we obtain the modified equations of motion 
for $i=\nu=0$:
\begin{equation} \label{gpfe1}
16\pi\rho = 2Tf_T-f ,
\end{equation}
for $i=0,\ \nu=a$:
\begin{equation} \label{gpfe2}
16\pi p = \left(\frac{4\bt'}{\bt r} - 2T\right) f_T+f ,
\end{equation}
and for $i=a,\ \nu=a$:
\begin{eqnarray} \label{gpfe3}
48\pi p &=& \left( \fr{4\bt'}{\bt} -\frac{2\alpha}{r}-rT\right) T' f_{TT}
- \frac{1}{\beta}\frac{d}{dr} (\beta rT - 4\beta')f_T \nonumber \\
&& + \left[ \frac{2\alpha}{r}\left( \fr{\bt'}{\bt} +\frac{4\bt'}{\alpha \bt}\right) - 5T\right] f_T + 3f . 
\end{eqnarray}
There is again one more non-vanishing component
of field equation: for $i=a,\ \nu=0$ we have
\begin{equation} \label{gpfe4}
\fr{d}{dr}(\fr{f_T}{\bt})=0 .
\end{equation}

This last equation quickly integrates, with $c$ being a constant, to 
\begin{equation} \label{fTb}
f_T (r) = c \beta (r)\, ,
\end{equation} 
which cannot be integrated further, because we do not know the functional forms of neither $T(r)$ nor $\beta(r)$.\footnote{Note that here and later in equation (\ref{lin}) we ignore an additive numerical constant, which corresponds to the cosmological constant.} Now we want to get a solution to the vacuum field equations exterior of the star which will be matched with the possible interior solution. From equation (\ref{gpfe1}), after setting $\rho =0$, and then using equation (\ref{fTb}), we find
\begin{equation} 
f(T) = 2c\beta (T) T
\end{equation}
Substituting this result into equation (\ref{gpfe2}), after setting $p =0$, we further find that
\begin{equation} 
\beta' = 0\, ,
\end{equation}
which implies that $\beta (r) = b = constant$.
Then (\ref{fTb}) can be integrated after all:
\begin{equation} \label{lin}
f(T) = 2cbT\, ,
\end{equation}  
with $c$ and $b$ constants.   

Thus we, or the freely falling observer, again find that $f(T)$ must be a linear function of $T$.
Hence we reach the same conclusion as in the previous section. Only possible form
of $f(T)$ function is the linear one and therefore other than the teleparallel equivalent of general
relativity no $f(T)$ theory of gravity is possible in the case of spherically symmetric neutron star problem.


\section{Conclusions}

In this paper, we aimed to analyze the $f(T)$ theories of gravity in
the strong field regime. $f(T)$ gravity has second order field
equations compared to fourth order field equations of $f(R)$ gravity.
Therefore it is an easier theory to study. However, it is shown in \cite{Sot1,Sot2} that the action of
$f(T)$ gravity is not Lorentz invariant. This means that the gravitational effects in $f(T)$ theories
are frame dependent and different choices of frame will result in different forms of field equations.
This strong background dependence of field equations of $f(T)$ gravity is
also discussed in \cite{Wang,Ferraro3}.

In the search of spherically symmetric relativistic star solution in $f(T)$ gravity
we found that the field equations forces a linear functional form
for $f(T)$. In other words, relativistic star solutions are possible only in
the case of teleparallel equivalent of general relativity. \emph{We reached to the same conclusion in two different physical reference frames:} one is the reference frame of a distant Schwarzschild observer (with a diagonal metric) and the other is the reference frame of a freely falling Gullstrand-Painlev\'e observer (with a non-diagonal metric).
This conclusion of the present paper points out to a serious problem.
Even though $f(T)$ theories of gravity might explain late time acceleration of
the universe \cite{Bengochea,Linder,Myrzakulov,Yang},  they cannot be a viable
theory of gravity due to unsatisfactory behavior in the strong field regime.
This paper demonstrates the importance of testing alternative theories of
gravity by neutron star physics. Tests in the strong field regime complement
the solar system and the cosmological tests which are done to decide if a particular
theory of gravity could be a candidate of a viable theory of reality.

Lastly, we should mention that a locally Lorentz invariant version of $f(T)$ gravity is proposed in \cite{Sot1,Sot2} and its implications for cosmology is investigated in \cite{Sot3}. Investigating the neutron star solutions in such a theory would be an interesting problem.


\section*{Acknowledgments}

We would like to thank K. Yavuz Ek\c{s}i and A. Sava\c{s} Arapo\u{g}lu
for helpful discussions.
C.D. is supported in part by the Turkish Council of Research and
Technology (T\"{U}B\.{I}TAK) through grant number 108T686.


\end{document}